\documentclass[journal=apchd5,manuscript=article]{achemso}

\usepackage{lineno}

\usepackage{amsmath,amssymb,bm}
\usepackage{physics}

\title{Trajectory probing of complex-frequency scattering with chirped analytic pulses}

\author{Alex Krasnok}
\email{akrasnok@fiu.edu}
\affiliation[Florida International University]
{Department of Electrical and Computer Engineering, Florida International University, Miami, Florida 33174, USA}

\author{Denis Seletskiy}
\affiliation[University of New Mexico]
{femtoQ Laboratory, Department of Physics and Astronomy, MSC07 4220, 1 University of New Mexico, Albuquerque NM, 87131-0001, USA}

\keywords{complex-frequency scattering, chirped analytic pulses, resonant scattering, analytic signals, contour probing, temporal coupled-mode theory}

\begin{document}


\begin{abstract}
Characterizing resonant scatterers is challenging because their poles and zeros usually lie away from the real-frequency axis, whereas most measurements sample only real frequencies and infer off-axis behavior from fitted models. Here we introduce complex-frequency chirped pulses: finite-energy analytic waveforms that probe a device continuously along a prescribed contour in the complex-frequency plane. We give a direct synthesis rule for an in-phase/quadrature (I/Q) waveform and show that finite-duration windowing deterministically distorts the realized trajectory, which makes it necessary to analyze only a central time interval where the window contribution is small. For stable linear time-invariant devices, we extract a time-local least-squares input--output ratio and identify when it follows the continued complex-frequency response, with errors that grow at higher traversal speeds and near resonant poles. Numerical tests on a coupled-mode resonator validate the method and show that closed contours enable an integer phase-winding consistency check. We also outline an implementation based on standard arbitrary waveform generation, I/Q modulation, coherent reception, and digital signal processing.
\end{abstract}

\section{Introduction}

 {Poles and zeros of the scattering response determine how resonant structures interact with waves. Poles encode complex modal frequencies and decay rates, while zeros encode destructive-interference conditions such as reflection cancellation and coherent-absorption channels. Because these singularities usually lie away from the real-frequency axis, interpreting measured spectra often requires extending a model into the complex-frequency plane.}

 {Most experiments sample only real frequencies and infer off-axis structure from fitted or approximated models, for example by vector fitting or AAA-based rational approximation over a finite band \cite{Gustavsen1999VF,Nakatsukasa2018AAA,Betz2024LPR_AAA,Binkowski2024PRB}. These methods can be highly accurate, but the off-axis continuation is constrained mainly by agreement on the real axis. As a result, discrepancies between models are often most important near poles and zeros.}

 {A complementary route is to shape the excitation directly in time. When the phase and envelope of an analytic input are varied in a controlled way, a stable linear time-invariant (LTI) device can be driven so that its output is well approximated by the continued response evaluated at a complex argument $\tilde{\omega}=\omega_r+i\omega_i$. Previous work has mainly targeted isolated complex-frequency points, enabling effects such as coherent virtual absorption and virtual critical coupling \cite{Baranov2017Optica,Radi2020ACSP}. Recent literature has cast these protocols as an operational form of complex-frequency continuation for wave systems \cite{KimScience2025}.}

 {Point-by-point excitation is not ideal when the goal is to test a fitted model over a neighborhood of a singularity or to probe how the response changes along a finite path. A single off-axis sample provides limited geometric information. In addition, practical implementations must use finite-energy waveforms, and the window that enforces finite duration also perturbs the realized complex frequency when the imaginary-frequency component is encoded through envelope growth or decay.}

 {Here we introduce complex-frequency chirped pulses as contour probes. Rather than reconstructing the full complex plane from a fitted model, we directly interrogate the response along a chosen contour. We prescribe a time-dependent instantaneous complex frequency $\tilde{\omega}(t)=\omega_r(t)+i\omega_i(t)$ that traces a contour $\mathcal{C}$ in the complex-frequency plane. We give a direct synthesis rule for the in-phase/quadrature (I/Q) baseband waveform, where $I$ and $Q$ are the real and imaginary components of the complex envelope. We show that finite windowing adds the deterministic correction $i\,d[\ln w(t)]/dt$, where $w(t)$ is the applied finite-duration window. This leads to a simple design rule: trajectory comparison and response extraction must be restricted to a central time interval where the window contribution is negligible. We then compare a time-local least-squares estimate of the input--output scattering ratio to $S(\tilde{\omega}(t))$ evaluated along the measured trajectory, derive a compact scaling estimate for the main error sources, and validate the method on a coupled-mode resonator. For closed contours, the extracted ratio also supports an integer phase-winding consistency check.}

\begin{figure}[t]
\centering\includegraphics[width=16cm]{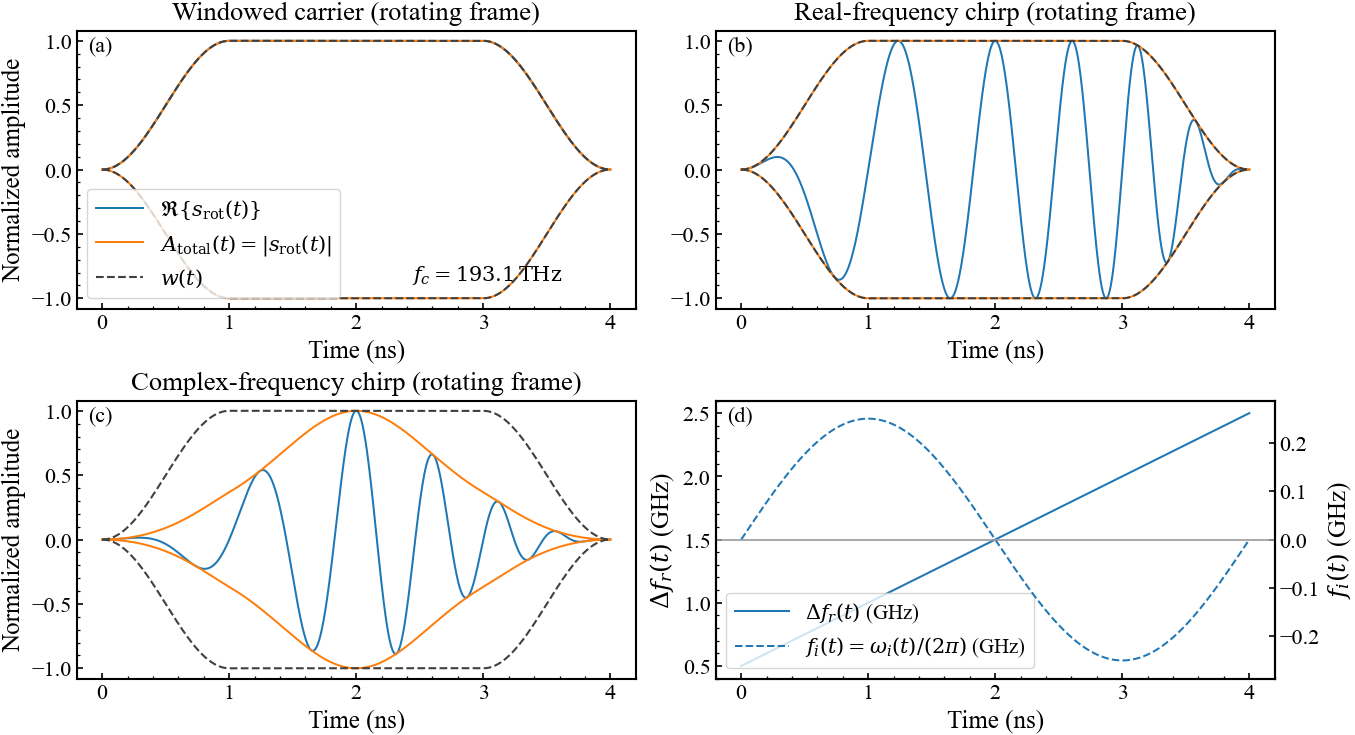}
\caption{ {\textbf{Ordinary chirps and complex-frequency chirped pulses in a rotating frame.}
Real part of the rotating-frame signal $s_{\mathrm{rot}}(t)=s_{\mathrm{in}}(t)e^{+i\omega_c t}$ (blue), its envelope $A_{\mathrm{total}}(t)=|s_{\mathrm{rot}}(t)|$ (orange), and the finite-duration window $w(t)$ (black dashed) for (a) a windowed carrier, (b) a real-frequency chirp defined by a programmed $\omega_r(t)$ with an approximately flat envelope over the central analysis interval, and (c) a complex-frequency chirp with the same $\omega_r(t)$ as in (b) but with a prescribed $\omega_i(t)$ that reshapes the envelope throughout the burst.
Panel (d) shows the programmed laws used for (b,c): real-frequency detuning $\Delta f_r(t)\equiv[\omega_r(t)-\omega_c]/(2\pi)$ (solid, left axis) and imaginary frequency $f_i(t)\equiv\omega_i(t)/(2\pi)$ (dashed, right axis).}}
\label{fig:Fig1n}
\end{figure}

\section{Complex-frequency chirped pulses and trajectories}

\subsection{Analytic signals and instantaneous complex frequency}

 {We use the harmonic convention $\exp(-i\omega t)$. For a complex exponential with $\omega=\omega_r+i\omega_i$, the magnitude scales as $\exp(\omega_i t)$, so $\omega_i>0$ corresponds to temporal growth and $\omega_i<0$ to temporal decay. We represent measured fields by analytic signals, obtained either by coherent I/Q reception or by forming the analytic associate of a real record \cite{Picinbono1997}.}

 {At a calibrated input reference plane---chosen after the transmitter chain and before the device under test so that the incident analytic field can be monitored directly---we write the analytic input as}
\begin{equation}
s_{\mathrm{in}}(t)=A(t)\,e^{-i\phi(t)},\qquad A(t)>0,
\end{equation}
 {and define the instantaneous complex angular frequency}
\begin{equation}
\tilde{\omega}(t)\equiv i\frac{d}{dt}\ln s_{\mathrm{in}}(t)
=\dot{\phi}(t)+i\frac{d}{dt}\ln A(t)
\equiv \omega_r(t)+i\omega_i(t).
\label{eq:inst_complex_freq}
\end{equation}
 {Here $\omega_r(t)$ has units of rad/s and gives the local phase slope, while $\omega_i(t)$ has units of s$^{-1}$ and gives the local logarithmic slope of the envelope. Equation~\eqref{eq:inst_complex_freq} implies
$A(t)=A(t_0)\exp\!\left(\int_{t_0}^{t}\omega_i(\tau)\,d\tau\right)$, so prescribing $\omega_i(t)$ is equivalent to prescribing the local exponential growth or decay rate of the analytic envelope.}

 {The quantity $\tilde{\omega}(t)$ is useful only when the analytic signal is effectively monocomponent and stays away from amplitude zeros. Near a null, differentiation of $\ln s_{\mathrm{in}}(t)$ becomes noise-sensitive and numerically unstable \cite{Boashash1992I,Boashash1992II}. In this work $\tilde{\omega}(t)$ plays two roles: it is the programmed target, and it is also a measured diagnostic. In data processing we estimate it from short-window fits to the unwrapped phase and log-amplitude of $s_{\mathrm{in}}(\tau)$ rather than by pointwise differentiation. The estimator and its discrete-time implementation are given in Supplement~1, Sec.~S1.}

 {Figure~\ref{fig:Fig1n} fixes the signal conventions and separates ordinary real-frequency chirps from complex-frequency chirped pulses. We choose a carrier at angular frequency $\omega_c$ and define the rotating-frame signal
$s_{\mathrm{rot}}(t)\equiv s_{\mathrm{in}}(t)e^{+i\omega_c t}$.
In coherent generation and reception, $s_{\mathrm{rot}}(t)$ is the complex baseband signal produced and detected through in-phase/quadrature channels. Figure~\ref{fig:Fig1n}(a) shows a windowed carrier. In the rotating frame the waveform is real up to a constant phase, so its envelope follows the window $w(t)$, and any nonzero $\omega_i(t)$ arises mainly from the window edges through $d[\ln w(t)]/dt$. Figure~\ref{fig:Fig1n}(b) shows a conventional real-frequency chirp defined by $\omega_r(t)$ and a window $w(t)$. Over a central interval where $w(t)\approx 1$ and varies slowly, the envelope is nearly constant, so $\omega_i(t)$ remains small and $\tilde{\omega}(t)$ stays close to the real axis. Figure~\ref{fig:Fig1n}(c) uses the same $\omega_r(t)$ as in (b) but prescribes a nonzero $\omega_i(t)$ over the burst. The envelope is then reshaped throughout the central interval, not only at the edges. This is the key practical difference used later: two waveforms can share the same real-frequency chirp while tracing different complex-frequency trajectories.}

\subsection{Trajectory synthesis and LTI tracking}

 {Trajectory synthesis follows directly from Eq.~\eqref{eq:inst_complex_freq}. Given a target instantaneous complex frequency
$\tilde{\omega}_{\mathrm{des}}(t)=\omega_r(t)+i\omega_i(t)$ on $t\in[0,T]$, an ideal infinite-energy analytic waveform that realizes this trajectory is}
\begin{equation}
s_{\mathrm{core}}(t)=s_0
\exp\!\left(\int_{0}^{t}\omega_i(\tau)\,d\tau\right)
\exp\!\left[-i\int_{0}^{t}\omega_r(\tau)\,d\tau\right],
\label{eq:core_waveform}
\end{equation}
 {with complex scale $s_0$. Finite energy is enforced by a window,}
\begin{equation}
s_{\mathrm{in}}(t)=w(t)\,s_{\mathrm{core}}(t), \qquad 0\le w(t)\le 1,
\label{eq:windowed_waveform}
\end{equation}
 {which deterministically deforms the realized instantaneous complex frequency:}
\begin{equation}
\tilde{\omega}(t)=\tilde{\omega}_{\mathrm{des}}(t)+i\frac{d}{dt}\ln w(t).
\label{eq:window_correction}
\end{equation}

 {This correction fixes the main analysis rule. Trajectory comparison and response extraction are restricted to a central interval
$[t_1,t_2]\subset(0,T)$ where the window-induced term is small. A convenient criterion is}
\begin{equation}
\left|\frac{d}{dt}\ln w(t)\right|\le \varepsilon_w,\qquad t\in[t_1,t_2],
\label{eq:flat_window_criterion}
\end{equation}
 {where $\varepsilon_w$ is chosen to be small compared with the characteristic scale of the intended $\omega_i(t)$ on the contour and with the required measurement accuracy. This form remains well defined even when the contour crosses the real axis and $\omega_i(t)=0$. Practical selection of $[t_1,t_2]$ in discrete time is described in Supplement~1, Sec.~S1.}

 {Closed contours impose a simple envelope constraint. Since
$|s_{\mathrm{core}}(t)|\propto\exp\!\left(\int_0^t\omega_i(\tau)\,d\tau\right)$, choosing}
\begin{equation}
\int_{0}^{T}\omega_i(t)\,dt=0
\label{eq:net_gain_zero}
\end{equation}
 {returns the unwindowed envelope to its initial level after one traversal. The required dynamic range is set by the extrema of
$\int_0^t\omega_i(\tau)\,d\tau$ within $[t_1,t_2]$ and must stay within the linear range of the modulator, detector, and analog-to-digital converter (ADC).}

\begin{figure}[t]
\centering\includegraphics[width=16cm]{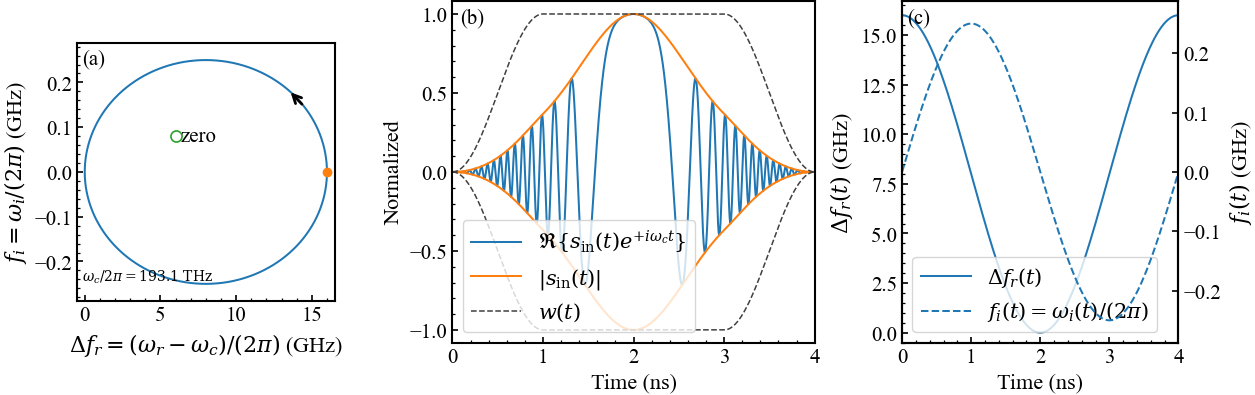}
\caption{ {\textbf{A closed complex-frequency trajectory and its finite-energy waveform.}
(a) Prescribed loop $\mathcal{C}$ traced by $\tilde{\omega}(t)=\omega_r(t)+i\omega_i(t)$, plotted as detuning $\Delta f_r(t)\equiv[\omega_r(t)-\omega_c]/(2\pi)$ and imaginary frequency $f_i(t)\equiv \omega_i(t)/(2\pi)$ (both in GHz). The orange marker indicates the start/end point on the real axis ($f_i=0$), the arrow indicates the traversal direction, and the ``zero'' marker denotes an example scattering zero enclosed by the loop. (b) Corresponding rotating-frame input waveform $\Re\{s_{\mathrm{in}}(t)e^{+i\omega_c t}\}$ (blue), envelope $|s_{\mathrm{in}}(t)|$ (orange), and finite-duration window $w(t)$ (black dashed). The central analysis interval is chosen from the flat part of $w(t)$. (c) Programmed laws $\Delta f_r(t)$ (solid, left axis) and $f_i(t)$ (dashed, right axis) used to realize the loop over one traversal.}}
\label{fig:cf_waveform}
\end{figure}

 {Figure~\ref{fig:cf_waveform} connects a target contour to a finite-duration waveform. Panel (a) shows the loop $\mathcal{C}$ traced by $\tilde{\omega}(t)$ in the plane of real detuning and imaginary frequency, where
$\Delta f_r(t)\equiv[\omega_r(t)-\omega_c]/(2\pi)$ and $f_i(t)\equiv\omega_i(t)/(2\pi)$.
Panel (c) shows one time-parameterization of the same loop through $\Delta f_r(t)$ and $f_i(t)$. In this example $f_i(t)$ has zero net area, consistent with Eq.~\eqref{eq:net_gain_zero}, so the core envelope returns to its initial level after the loop. Panel (b) shows the corresponding time-domain waveform and its envelope. The flat portion of $w(t)$ identifies the interval where Eq.~\eqref{eq:flat_window_criterion} holds and where trajectory comparisons are performed.}

 {In a coherent transmitter, the waveform is implemented as a complex baseband envelope on a carrier at $\omega_c$. Here $I(t)$ and $Q(t)$ are the in-phase and quadrature components of that baseband signal. Writing
$\Delta\omega_r(t)=\omega_r(t)-\omega_c$, the designed baseband drive is}
\begin{equation}
s_{\mathrm{bb}}(t)=I(t)+iQ(t)
=w(t)\exp\!\left(\int_{0}^{t}\omega_i(\tau)\,d\tau\right)
\exp\!\left[-i\int_{0}^{t}\Delta\omega_r(\tau)\,d\tau\right].
\label{eq:baseband_design}
\end{equation}
 {Analog filtering and I/Q imbalance distort $I(t)$ and $Q(t)$, so trajectory probing treats $\tilde{\omega}(t)$ as a measured quantity rather than an assumed one. It is estimated from the monitored input $s_{\mathrm{in}}(t)$ over $[t_1,t_2]$ and compared with the designed contour. Calibration and digital pre-distortion strategies are discussed in Supplement~1, Sec.~S5.}

 {The connection between a time-varying complex-frequency drive and the continued response is clearest for stable LTI models. Consider}
\begin{align}
\dot{\bm{a}}(t) &= \bm{A}\bm{a}(t)+\bm{B}\bm{s}_{\mathrm{in}}(t), \\
\bm{s}_{\mathrm{out}}(t) &= \bm{C}\bm{a}(t)+\bm{D}\bm{s}_{\mathrm{in}}(t),
\end{align}
 {with $\Re\lambda_k(\bm{A})<0$. Under the $\exp(-i\omega t)$ convention, the transfer function}
\begin{equation}
\bm{S}(\omega)=\bm{D}+\bm{C}\left(-i\omega\bm{I}-\bm{A}\right)^{-1}\bm{B}
\end{equation}
 {has no poles for $\Im\{\omega\}>0$ and is analytic in the upper half-plane. Away from poles, the same meromorphic expression can be evaluated at complex frequencies elsewhere in the plane. To expose the tracking condition, we factor the input into a rapidly varying exponential that carries the programmed complex frequency and a slowly varying prefactor:
$\bm{s}_{\mathrm{in}}(t)=\bm{u}(t)\exp\!\left[-i\int_{0}^{t}\tilde{\omega}(\tau)\,d\tau\right]$.
Here $\bm{u}(t)$ collects any residual slow amplitude variation not included in the nominal contour. If $\bm{u}(t)$ and $\tilde{\omega}(t)$ vary slowly compared with the internal relaxation, the output is approximately quasi-steady:}
\begin{equation}
\bm{s}_{\mathrm{out}}(t)\approx \bm{S}\!\left(\tilde{\omega}(t)\right)\bm{s}_{\mathrm{in}}(t).
\label{eq:instantaneous_mapping}
\end{equation}

 {To test Eq.~\eqref{eq:instantaneous_mapping} from time records, we estimate a time-local complex ratio. For a single input and output we use a weighted least-squares estimator with a real, nonnegative window $g(\tau-t)$:}
\begin{equation}
\hat{r}(t)=
\frac{\int s_{\mathrm{out}}(\tau)\,s_{\mathrm{in}}^\ast(\tau)\,g(\tau-t)\,d\tau}
{\int |s_{\mathrm{in}}(\tau)|^2\,g(\tau-t)\,d\tau},
\label{eq:ls_ratio}
\end{equation}
 {and we report $\hat{r}(t)$ only when the denominator exceeds a noise-dependent threshold. A discrete-time implementation and recommended window choices are given in Supplement~1, Sec.~S1. The same idea extends to multiport measurements as a local matrix least-squares estimate of $\bm{S}(t)$, provided the driven inputs span the relevant subspace over the local fit window; see Supplement~1, Sec.~S6.}

 {To summarize the limits of quasi-steady tracking, define the mismatch
$\bm{e}(t)\equiv \bm{s}_{\mathrm{out}}(t)-\bm{S}(\tilde{\omega}(t))\bm{s}_{\mathrm{in}}(t)$ and}
\begin{equation}
\alpha(t)\equiv \sigma_{\min}\!\left(\bm{A}+i\tilde{\omega}(t)\bm{I}\right),
\qquad
\left\|\left[\bm{A}+i\tilde{\omega}(t)\bm{I}\right]^{-1}\right\|=\frac{1}{\alpha(t)}.
\end{equation}
 {When $\bm{u}(t)$ and $\tilde{\omega}(t)$ are smooth and vary on a timescale slow compared to $1/\alpha(t)$, an adiabatic expansion yields the scaling estimate}
\begin{equation}
\|\bm{e}(t)\|\;\lesssim\; \frac{c_1\,\|\dot{\bm{u}}(t)\|}{\alpha(t)}
+\frac{c_2\,|\dot{\tilde{\omega}}(t)|\,\|\bm{u}(t)\|}{\alpha^2(t)},
\label{eq:tracking_scaling}
\end{equation}
 {where the constants $c_1$ and $c_2$ depend on $\|\bm{B}\|$, $\|\bm{C}\|$, and the chosen norms. Equation~\eqref{eq:tracking_scaling} highlights two coupled failure modes: faster traversal increases $|\dot{\tilde{\omega}}(t)|$, and proximity to poles decreases $\alpha(t)$. Both effects increase the tracking error. Supplement~1, Sec.~S2 derives the corresponding bound for the single-mode coupled-mode model used in the numerical example below.}

\begin{figure}[t]
\centering\includegraphics[width=14cm]{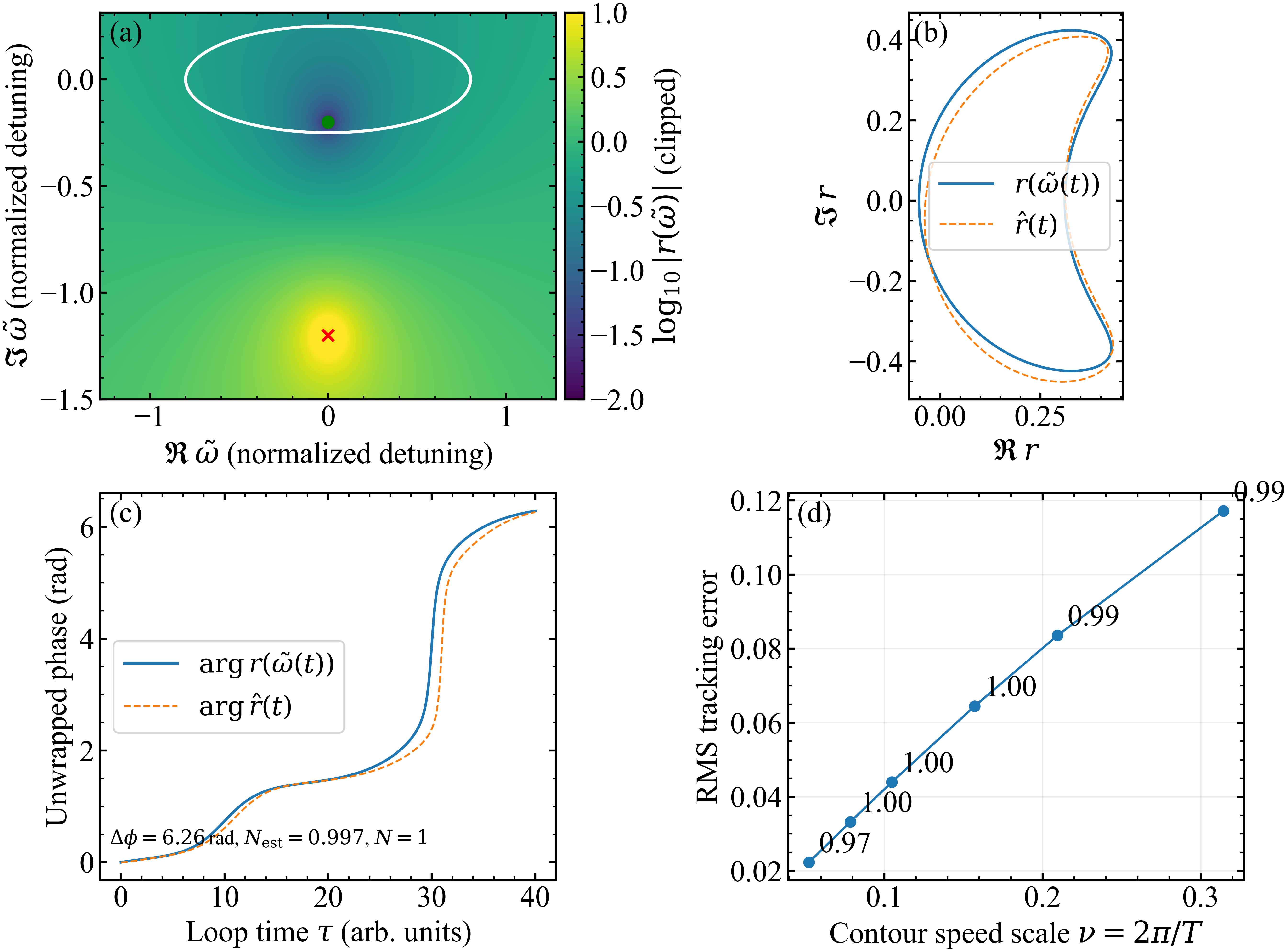}
\caption{\textbf{Numerical validation of trajectory probing and its speed dependence.}
(a) Map of $\log_{10}|r(\tilde{\omega})|$ for a one-port resonator in normalized detuning units. The contour $\mathcal{C}$ (white) encloses a zero (green) while avoiding a pole (red). (b) Complex-$r$ trajectory for a baseline loop with $T=40$ arb.\ units and $\nu=2\pi/T\simeq0.157$: continued response $r(\tilde{\omega}(t))$ (solid) and time-local least-squares estimate $\hat{r}(t)$ (dashed). (c) Unwrapped phase for the same loop, giving $\Delta\phi=6.26$ rad, $N_{\mathrm{est}}=\Delta\phi/(2\pi)=0.997$, and $N=1$. (d) RMS tracking error versus traversal speed for fixed contour geometry, evaluated over the central interval; annotations show $N_{\mathrm{est}}$, illustrating increased mismatch and small departures from integer-$2\pi$ phase accumulation at higher speed.}\label{fig:validation}
\end{figure}

 {Figure~\ref{fig:validation} validates trajectory probing on a one-port resonator described by temporal coupled-mode theory (TCMT) \cite{Haus1991,Fan2003TCMT}. For the $\exp(-i\omega t)$ convention, a standard one-pole/one-zero reflection coefficient is}
\begin{equation}
r(\tilde{\omega})=
\frac{-i(\tilde{\omega}-\omega_0)+(\gamma_0-\gamma_c)}
{-i(\tilde{\omega}-\omega_0)+(\gamma_0+\gamma_c)},
\label{eq:tcmt_r}
\end{equation}
 {where $\omega_0$ is the resonant angular frequency (rad/s), $\gamma_0$ is the internal loss rate (s$^{-1}$), and $\gamma_c$ is the external coupling rate (s$^{-1}$). The pole is at $\tilde{\omega}_p=\omega_0-i(\gamma_0+\gamma_c)$, while the zero is at $\tilde{\omega}_z=\omega_0-i(\gamma_0-\gamma_c)$. Numerical parameters and normalization are given in Supplement~1, Sec.~S4.}

 {Figure~\ref{fig:validation}(a) shows $\log_{10}|r(\tilde{\omega})|$ and the chosen contour $\mathcal{C}$. The contour encloses the zero while staying away from the pole. This separation sets the tracking margin because $\alpha(t)$ decreases near poles, consistent with Eq.~\eqref{eq:tracking_scaling}. Figure~\ref{fig:validation}(b) compares the continued response sampled along the realized trajectory, $r(\tilde{\omega}(t))$, with the time-local estimate $\hat{r}(t)$ extracted from simulated time records using Eq.~\eqref{eq:ls_ratio}. Their agreement is the direct check that the waveform probes the intended region and that the response remains close to quasi-steady along the traversal.}

 {To vary traversal speed without changing the contour geometry, we parameterize the contour by $u\in[0,2\pi]$ and traverse it with
$u(t)=2\pi t/T$. The contour-speed scale is then}
\begin{equation}
\nu\equiv \dot{u}=\frac{2\pi}{T},
\label{eq:contour_speed_def}
\end{equation}
 {so that $d\tilde{\omega}/dt=(d\tilde{\omega}/du)\,\nu$. Faster traversal increases $|\dot{\tilde{\omega}}|$ while leaving $\tilde{\omega}(u)$ unchanged, isolating the speed dependence predicted by Eq.~\eqref{eq:tracking_scaling}.}

 {Figure~\ref{fig:validation}(c) reports the unwrapped phases of $r(\tilde{\omega}(t))$ and $\hat{r}(t)$ along one traversal. Unwrapping enforces continuity across $\pm\pi$ branch cuts (for example via the Itoh criterion) \cite{Itoh1982PhaseUnwrap}; implementation details are given in Supplement~1, Sec.~S3. The net phase accumulation is
$\Delta\phi=\arg\hat{r}(t_2)-\arg\hat{r}(t_1)$ after unwrapping, and we define the winding estimate}
\begin{equation}
N_{\mathrm{est}}\equiv \frac{\Delta\phi}{2\pi}.
\label{eq:Nest_def}
\end{equation}
 {Figure~\ref{fig:validation}(d) quantifies the speed dependence using the RMS tracking error}
\begin{equation}
\epsilon_{\mathrm{rms}}\equiv
\left[\frac{1}{t_2-t_1}\int_{t_1}^{t_2}\left|\hat{r}(t)-r\!\left(\tilde{\omega}(t)\right)\right|^2\,dt\right]^{1/2},
\label{eq:rms_error_def}
\end{equation}
 {evaluated over the central interval to suppress window-edge effects. A discrete-time version of Eq.~\eqref{eq:rms_error_def} and the simulation recipe are given in Supplement~1, Sec.~S4.}

 {Closed trajectories also permit a phase-winding consistency check. If tracking holds so that $\hat{r}(t)\approx r(\tilde{\omega}(t))$ and the contour does not cross singularities, then the argument principle gives}
\begin{equation}
\Delta_{\mathcal{C}}\arg r = 2\pi\left(N_z-N_p\right),
\end{equation}
 {where $N_z$ and $N_p$ are the numbers of enclosed zeros and poles counted with multiplicity \cite{Ahlfors1979}. In data we report $\hat{N}=\mathrm{round}(N_{\mathrm{est}})$ together with the residual}
\begin{equation}
\epsilon\equiv \left|\Delta\phi-2\pi\hat{N}\right|,
\end{equation}
 {which quantifies how closely the measured phase accumulation approaches an integer multiple of $2\pi$.}

 {The main limits of trajectory probing follow from Eqs.~\eqref{eq:window_correction} and \eqref{eq:tracking_scaling}. Windowing adds the deterministic correction $i\,d[\ln w(t)]/dt$, which is why analysis must be confined to a central interval. The programmed $\omega_i(t)$ sets the required amplitude dynamic range through the extrema of $\int \omega_i\,dt$, while $\Delta\omega_r(t)$ sets the required baseband bandwidth. Pole proximity reduces $\alpha(t)$ and increases memory effects, so contour placement and traversal speed must be chosen together. Finally, the method assumes linearity and time invariance over the measurement interval; repeating the same contour and comparing the recovered $\tilde{\omega}(t)$ and $\hat{r}(t)$ across bursts provides a direct check for drift or nonlinearity.}

\section{Implementation workflow}

\begin{figure}[t]
\centering
\includegraphics[width=16cm]{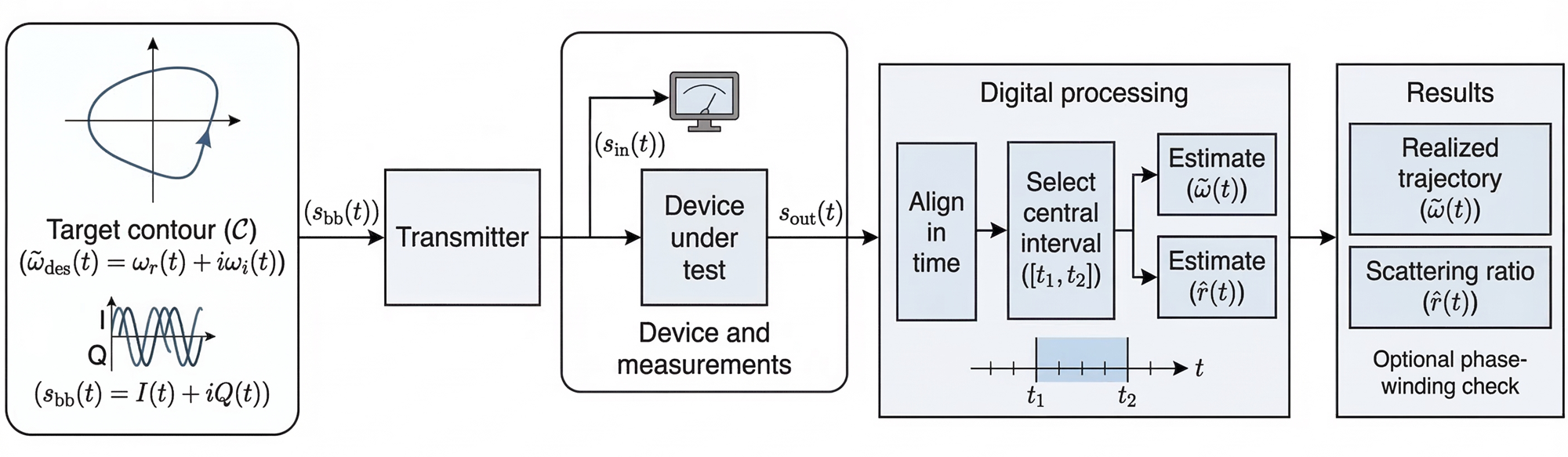}
\caption{\textbf{Implementation workflow for trajectory probing.}
A target contour $\mathcal{C}$ in the complex-frequency plane defines the desired instantaneous complex frequency $\tilde{\omega}_{\mathrm{des}}(t)=\omega_r(t)+i\omega_i(t)$ and the corresponding baseband waveform $s_{\mathrm{bb}}(t)=I(t)+iQ(t)$. This waveform drives the transmitter, which launches the signal toward the device under test. The incident signal is monitored at the input, giving $s_{\mathrm{in}}(t)$, while the device response is recorded as $s_{\mathrm{out}}(t)$. Digital processing then aligns the input and output records in time, selects a central interval $[t_1,t_2]$ where window-induced distortion is small, estimates the realized complex-frequency trajectory $\tilde{\omega}(t)$ from the monitored input, and extracts the time-local scattering ratio $\hat{r}(t)$ from the synchronized records. For closed contours, the recovered ratio can also be used for an optional phase-winding check.}
\label{fig:schematic}
\end{figure}

 {Trajectory probing is implemented as a finite-duration burst measurement that produces two synchronized complex analytic records: the incident field $s_{\mathrm{in}}(t)$ at the input reference plane and the corresponding output field $s_{\mathrm{out}}(t)$ at the measured port. The goal is to realize a prescribed contour $\mathcal{C}$ in the complex-frequency plane, verify the realized trajectory from the monitored input, and extract a time-local scattering ratio along a central analysis interval; see Fig.~\ref{fig:schematic}.}

 {Waveform design begins with the target instantaneous complex angular frequency $\tilde{\omega}_{\mathrm{des}}(t)=\omega_r(t)+i\omega_i(t)$ under the $\exp(-i\omega t)$ convention, where $\omega_r(t)$ sets the local phase slope and $\omega_i(t)$ sets the local exponential growth or decay rate of the analytic envelope. A finite-energy drive is then formed by applying a window $w(t)$ to the ideal trajectory-realizing core waveform, as in Eqs.~\eqref{eq:core_waveform}--\eqref{eq:windowed_waveform}. The chosen contour must satisfy three practical limits: the extrema of $\int \omega_i\,dt$ must fit within the linear amplitude range of the modulation and detection chain, the variation of $\Delta\omega_r(t)$ must fit within the available baseband bandwidth, and the monitored input must remain above the receiver noise floor over the central interval.}

 {In a coherent transmitter the waveform is implemented as a complex baseband signal $s_{\mathrm{bb}}(t)=I(t)+iQ(t)$ on a carrier at angular frequency $\omega_c$, using Eq.~\eqref{eq:baseband_design}. The same structure applies in optical and microwave settings: a digital waveform source generates $I(t)$ and $Q(t)$, which drive an I/Q modulator that upconverts the baseband to the carrier \cite{Kawanishi2011ParallelMM}. The protocol does not depend on a specific modulator architecture as long as the transmitted field can be monitored and the received field can be coherently recovered as a complex analytic record. Hardware impairments such as finite bandwidth, I/Q imbalance, and group-delay ripple are summarized in Supplement~1, Sec.~S5.}

 {Trajectory probing requires an input monitor. The realized contour is extracted from the measured $s_{\mathrm{in}}(t)$ rather than assumed from the programmed waveform, so the input is recorded after the transmitter chain at the same calibrated reference plane used for delay alignment and channel calibration. This monitor supplies both the signal used to estimate $\tilde{\omega}(t)$ and the reference used in the local ratio estimator, which reduces sensitivity to slow gain drift and envelope variation.}

 {Before forming any local ratio, the input and output records must be aligned in time. Let $\tau_d$ denote the relative delay between the measured $s_{\mathrm{in}}(t)$ and $s_{\mathrm{out}}(t)$ referenced to the chosen planes. The delay can be estimated from calibration data or from the burst itself by cross-correlation and then compensated so that the two records correspond to the same interaction time in the device. Generalized cross-correlation is useful when the noise is colored or when the monitor and output paths have different filtering \cite{KnappCarter1976}. A discrete-time alignment procedure is given in Supplement~1, Sec.~S1.}

 {After alignment, analysis is restricted to a central interval $[t_1,t_2]$ where the window slope is small, transients from the burst edges have died away, and the monitored input magnitude remains above a noise floor. Over this interval we estimate the local scattering ratio with Eq.~\eqref{eq:ls_ratio}. The same windowed least-squares form generalizes to multiport measurements when repeated bursts or orthogonal multiport excitations provide enough independent input combinations over the fit window; see Supplement~1, Sec.~S6.}

 {Over the same interval we estimate $\tilde{\omega}(t)=i\,d[\ln s_{\mathrm{in}}(t)]/dt$ from short-window fits to the input record rather than by pointwise differentiation, which reduces noise sensitivity and avoids numerical instability near amplitude nulls \cite{Boashash1992I,Boashash1992II}. In practice, comparing the measured $\tilde{\omega}(t)$ with the designed $\tilde{\omega}_{\mathrm{des}}(t)$ is the main check that the intended contour is actually traced after transmitter filtering and receiver processing. The estimator and its discrete-time implementation are given in Supplement~1, Sec.~S1.}

 {For closed contours, the unwrapped phase of $\hat{r}(t)$ provides an additional consistency check. From the net phase accumulation $\Delta\phi$ over $[t_1,t_2]$ we form $N_{\mathrm{est}}=\Delta\phi/(2\pi)$ and report both the rounded winding $\hat{N}=\mathrm{round}(N_{\mathrm{est}})$ and the residual $\epsilon=\left|\Delta\phi-2\pi\hat{N}\right|$. Small residuals support the interpretation of the measured loop in terms of enclosed zeros and poles. Phase-unwrapping details and robustness metrics are summarized in Supplement~1, Sec.~S3.}

 {This workflow makes the core assumptions visible in the data. The monitored input verifies the realized trajectory, the central interval isolates the part of the burst where the window correction is small, and repeated traversals provide a direct check for drift, nonlinearity, and time variance. Supplement~1 summarizes robustness metrics for reversal, contour deformation, and speed reparameterization.}

\section{Outlook}

 {Trajectory probing turns complex-frequency excitation into a contour-level measurement tool: one finite-energy burst samples $S(\tilde{\omega})$ along a prescribed path while the actual path is reconstructed from the measured input. The central design constraint is still Eq.~\eqref{eq:window_correction}. Because windowing deforms the trajectory in a deterministic way, useful measurements require a central interval where that correction is small and where the tracking condition remains satisfied.}

 {A natural extension is multiport trajectory probing. When coherent records are available for several driven and measured ports, a local matrix estimate of $\bm{S}(t)$ can be formed and compared with a fitted or simulated $\bm{S}(\tilde{\omega})$ along the measured trajectory. In practice this requires input waveforms that span the relevant port space, for example through repeated bursts or orthogonal multiport drives. Invariants such as $\arg\det\bm{S}$ or the phase winding of selected eigenvalues then provide closed-contour diagnostics that do not depend only on pointwise amplitude agreement.}

 {A second direction is model-guided contour design. Rational approximation methods such as vector fitting and AAA provide compact meromorphic models over finite bands \cite{Gustavsen1999VF,Nakatsukasa2018AAA,Betz2024LPR_AAA}. These models can be used to place contours with explicit margins to poles and to choose traversal schedules that respect the window constraint, the available dynamic range, the baseband bandwidth, and the tracking scaling in Eq.~\eqref{eq:tracking_scaling}. In this way, trajectory probing can be used to stress-test a fitted model precisely where its off-axis continuation is most sensitive.}

 {A third direction is closed-loop correction of trajectory distortion. Because $\tilde{\omega}(t)$ is estimated from the monitored input, calibration and digital pre-distortion can be used to reduce deformations caused by analog filtering and I/Q imperfections. Once the residual trajectory error is small compared with the contour size, the same measurement protocol can be used repeatedly to validate fitted models, probe neighborhoods of poles and zeros, and track how off-axis structure moves under controlled perturbations.}

\section{Conclusions}

 {We introduced complex-frequency chirped pulses as a trajectory-based way to probe scattering responses away from the real-frequency axis. A finite-energy analytic burst is designed to follow a prescribed contour, and the actual contour is reconstructed from the monitored input rather than assumed from the programmed waveform. Two constraints govern the method. First, finite windowing deterministically perturbs the trajectory, so response extraction must be confined to a central interval where the window contribution is small. Second, accurate probing requires quasi-steady tracking, so we estimate a time-local input--output ratio by least squares and compare it with the continued response evaluated along the measured trajectory. The resulting error grows with traversal speed and with proximity to poles. Numerical tests on a one-port coupled-mode resonator validate the approach and show that closed contours support an integer phase-winding check when tracking holds. Supplement~1 gives the discrete-time estimators, the coupled-mode tracking bound, implementation details, and additional robustness metrics.}

\section*{Notes}
The authors declare no conflicts of interest.

\begin{acknowledgement}
Alex Krasnok acknowledges financial support from the U.S. Department of Energy (DoE) and the U.S. Air Force Office of Scientific Research (AFOSR).
\end{acknowledgement}

\begin{suppinfo}
See Supplement~1 for supporting content. Supplement~1 provides implementation and validation details as follows: Sec.~S1 describes the discrete-time estimators for $\tilde{\omega}(t)$ and $\hat{r}(t)$, selection of the analysis interval, and delay alignment; Sec.~S2 derives the tracking bound for the single-mode coupled-mode model used in Fig.~\ref{fig:validation}; Sec.~S3 details phase unwrapping, winding extraction, and robustness metrics for traversal reversal, controlled contour deformation, and speed reparameterization; Sec.~S4 gives reproducible simulation parameters for Fig.~\ref{fig:validation}; Sec.~S5 discusses transmitter/receiver impairments, calibration, and iterative digital pre-distortion; Sec.~S6 presents a multiport extension and matrix-valued winding diagnostics.
\end{suppinfo}

\bibliography{refs}

\end{document}